\documentclass[12pt,prd,showkeys,superscriptaddress,tightenlines]{revtex4}
\usepackage{times}
\usepackage{amsmath}
\usepackage{graphicx}
\usepackage{epsfig}
\usepackage{dcolumn}
\usepackage{bm}
\usepackage{color}
\usepackage{longtable}
\usepackage{array}
\usepackage{multirow}
\usepackage{appendix}

\newcommand{\BR}{{\cal B}}

\newcommand{\piz}{\pi^0}
\newcommand{\etap}{\eta^{\prime}}
\newcommand{\hc}{h_c}

\newcommand{\etac}{\eta_c}
\newcommand{\etacp}{\eta_c(2S)}

\newcommand{\psp}{\psi(2S)}

\newcommand{\jpsi}{J/\psi}
\newcommand{\EE}{e^+e^-}

\newcommand{\GG}{\gamma\gamma}
\newcommand{\pp}{\pi^+\pi^-}
\newcommand{\ppp}{\pi^+\pi^-\pi^0}
\newcommand{\kk}{K^+K^-}

\newcommand{\ppb}{p\bar{p}}
\newcommand{\ks}{K_{S}^{0}}

\newcommand{\et}{\eta_c\to \gamma\gamma}
\newcommand{\eth}{B^+\to \eta_cK^+}
\newcommand{\eft}{\eta_c\to p\bar{p}}
\newcommand{\efi}{\eta_c\to K\bar{K}\pi}
\newcommand{\es}{\eta_c\to \kk\eta}
\newcommand{\ese}{\eta_c\to \pi^+\pi^-\etap}
\newcommand{\ee}{\eta_c\to \pi^+\pi^-p\bar{p}}
\newcommand{\en}{\eta_c\to \ks K^\pm\pi^\mp\pp}
\newcommand{\ete}{\eta_c\to K^+ K^-\pi^+\pi^-\pi^0}
\newcommand{\eel}{\eta_c\to \phi\phi}
\newcommand{\etw}{\eta_c\to \phi K^+K^-}
\newcommand{\etht}{\eta_c\to \pi^+\pi^-\eta}
\newcommand{\ef}{\eta_c\to 2(\pi^+\pi^-)}
\newcommand{\efit}{\eta_c\to K^+K^-\pi^+\pi^-}
\newcommand{\esit}{\eta_c\to 2(K^+K^-)}
\newcommand{\eset}{\eta_c\to 2(\pi^+\pi^-\pi^0)}

\newcommand{\epth}{B^+\to \eta_c(2S)K^+}
\newcommand{\epf}{\eta_c(2S)\to p\bar{p}}
\newcommand{\epfi}{\eta_c(2S)\to K\bar{K}\pi}
\newcommand{\eps}{\eta_c(2S)\to \kk\eta}
\newcommand{\epse}{\eta_c(2S)\to \pi^+\pi^-\eta'}
\newcommand{\epe}{\eta_c(2S)\to \pi^+\pi^-p\bar{p}}
\newcommand{\epn}{\eta_c(2S)\to \ks K^\pm\pi^\mp\pp}
\newcommand{\epte}{\eta_c(2S)\to K^+ K^-\pi^+\pi^-\pi^0}
\newcommand{\pse}{\psi(2S)\to\pi^0\hc\to \piz\gamma\eta_c}
\newcommand{\psep}{\psi(2S)\to \gamma\eta_c(2S)}
\newcommand{\efia}{\eta_c\to K^+K^-\pi^0}
\newcommand{\efib}{\eta_c\to \ks K^{\pm}\pi^{\mp}}
\newcommand{\epfia}{\eta_c(2S)\to K^+K^-\pi^0}
\newcommand{\epfib}{\eta_c(2S)\to \ks K^{\pm}\pi^{\mp}}
\newcommand{\etha}{B^0\to \eta_c\ks}
\newcommand{\ethb}{B^0\to \eta_cK^0}
\newcommand{\eptha}{B^0\to \eta_c(2S) \ks}

\newcommand{\bt}{BESIII}
\newcommand{\cl}{CLEO}
\newcommand{\bb}{BaBar}
\newcommand{\lh}{LHCb}
\newcommand{\be}{Belle}
\newcounter{myct}
\newcommand\myct{\stepcounter{myct}\arabic{myct}}

\renewcommand{\baselinestretch}{1.0}

\begin{document}

\title{New puzzle in charmonium decays}
\author{Hongpeng Wang}
 \email{wanghp@ihep.ac.cn}
\author{Chang-Zheng Yuan}
 \email{yuancz@ihep.ac.cn}
 \affiliation{Institute of High Energy Physics, Chinese Academy of Sciences,
 Beijing 100049, China}
 \affiliation{University of Chinese Academy of Sciences, Beijing 100049, China}

\date{\today}

\begin{abstract}

By analyzing existing data on pseudoscalar charmonium decays, we
obtain the ratio of the branching fractions of $\etacp$ and
$\etac$ decays into ten different final states with light hadrons.
For the first time, we test the two existing theoretical
predictions of these decays, and find that the experimental data
are significantly different from both of them. The lack of
observation of any decay mode with higher rate in $\etacp$ than in
$\etac$ decays suggests very unusual decay dynamics in
pseudoscalar charmonium decays to be identified. We also report
the first model-independent evaluation of the partial width of
$\etacp\to \GG$ ($2.21_{-0.64}^{+0.88}$~keV) and improved
determination of that of $\etac\to \GG$
($5.43_{-0.38}^{+0.41}$~keV). The latter shows a tension with the
most recent lattice QCD calculation.
\end{abstract}


\maketitle

Charmonium states are bound state of a charmed quark ($c$) and a
charmed antiquark ($\bar{c}$). Since the discovery of the first
charmonium state, the $\jpsi$, at BNL~\cite{ting} and at
SLAC~\cite{richter} in 1974, all the charmonium states below the
open-charm threshold and a few charmonium states above the
open-charm threshold have been established; the measured spectrum
of the states agrees well with theoretical calculations based on
QCD~\cite{Brambilla:2004jw,review4,Brambilla:2014jmp} and
QCD-inspired potential models~\cite{eichten,godfrey,barnes}. On
the contrary, the decays of the charmonium states into light
hadrons which must proceed via the annihilation of the charmed
quark-antiquark pair, are still poorly known, although they are
governed by the same QCD theory.

The first calculation by Appelquist and
Politzer~\cite{Appelquist:1974zd} using perturbative QCD related
the hadronic decays of $\jpsi$ and its radial excited sibling
$\psp$ to their leptonic decays, and predicted
\[
 Q^V =\frac{\BR(\psp\to hadrons)}{\BR(\jpsi\to hadrons)}
     =\frac{\BR(\psp\to \EE)}{\BR(\jpsi\to \EE)}.
\]
The ratio was found to be around 12\% using the branching
fractions of the leptonic decays at that time, and this was called
``12\% rule'' since then, although the most recent ratio is
$(13.3\pm 0.3)\%$~\cite{PDG}. Extending $Q^V$ of inclusive decays
of charmonium to light hadrons to each individual hadronic final
state, $h$, Mark II experiment tested $Q^V_h$ with eight final
states~\cite{Franklin:1983ve} and found two modes were severely
suppressed relative to 12\% while the other six modes agree with
12\% reasonably well, and the $\rho\pi$ mode was suppressed by
more than an order of magnitude so this was referred to as
``$\rho\pi$ puzzle''. Many theoretical explanations have been put
forth to decipher this puzzle~\cite{Mo:2006cy}, some attribute the
small $Q^V_{\rho\pi}$ to the enhanced branching fraction of
$\jpsi$ decays, some to the suppressed branching fraction of
$\psp$ decays, and some other to some dynamics which may affect
both $\psp$ and $\jpsi$ decays but in a different way. Improved
measurements from BES, CLEOc, and lately BESIII experiments
confirmed the Mark II observations and tested various theoretical
models~\cite{PDG}, it is found that none of these models can solve the
``$\rho\pi$ puzzle'' and all the newly available data
satisfactorily~\cite{Mo:2006cy}.

As the spin-partners of $\jpsi$ and $\psp$ respectively, the
spin-singlets $\etac$ and $\etacp$ may decay into light hadrons in
a similar way as their spin-triplets partners. Anselmino,
Genovese, and Predazzi assumed~\cite{Anselmino:1991es}
\[
 \frac{\BR(\etacp\to hadrons)}{\BR(\etac\to hadrons)}
 \approx \frac{\BR(\psp\to hadrons)}{\BR(\jpsi\to hadrons)} = Q^V,
\]
while Chao, Gu, and Tuan argued that~\cite{Chao:1996sf}
\[
 Q^P = \frac{\BR(\etacp\to hadrons)}{\BR(\etac\to hadrons)}
 \approx 1.
\]
These two predictions deviate by a factor of seven and
should be tested with experimental data.

The theoretical work was clearly ahead of time since the $\etacp$
was first observed in 2002~\cite{Belle:2002bnx} and until now only
three hadronic decays of it were listed with branching fractions
and the uncertainties are more than 50\%~\cite{PDG}.

By examining the experimental data available for $\etacp$ decays
(cited by the PDG~\cite{PDG} and those listed in the
Appendix~\ref{app}), we found an amazing fact that in most of the
cases both $\etac$ and $\etacp$ were measured in an experiment at
the same time, so this allows a very convenient way of determining
the ratio of the branching fractions and to test the theoretical
predictions. We scrutinize the experimental measurements and
select only the reliable results and do a global fit to extract
properties related to the $\etac$ and $\etacp$ states.

There are mainly three categories of measurements related to
$\etac$ and $\etacp$ states: two-photon processes ($\GG\to
\etac~(\etacp)$), $B$ meson decays ($B\to K\etac~(\etacp)$), and
charmonium decays ($\psp\to \gamma \etac~(\etacp)$, $\jpsi\to
\gamma \etac$, and $h_c\to \gamma \etac$). In many of the cases,
experimental measurements are the ratio or the product of the
branching fractions or partial widths. With the help of
measurements of a few absolute branching fractions and the total
widths of $\etac$ ($\Gamma_{\etac}$) and $\etacp$
($\Gamma_{\etacp}$), we are able to determine the branching
fractions of $\etac$ and $\etacp$ decays and the ratios, as well
as their partial widths to $\GG$ ($\Gamma_{\etac\to \GG}$ and
$\Gamma_{\etacp\to \GG}$).

The $\etac$-related measurements before 1995 were obtained by
using $\etac$ mass and width that are significantly smaller than
the recent results~\cite{PDG}. They are not used in our analysis
since the results are biased and the precision is low. The
$\etac$-related measurements from $\jpsi\to \gamma \etac$ were
biased by neglecting the interference between $\etac$ and
non-$\etac$ amplitudes and using unreliable line shape of $\etac$
resonance in this $M1$ transition~\cite{Segovia:2021bjb}. They are
also not used in our analysis.

We are left with 97 measurements from the AMY, BaBar, Belle,
BESIII, CLEO, DELPHI, E760, E835, and LHCb experiments, as listed
in Appendix~\ref{app}. We do a least-squares fit with 29
parameters, and the $\chi^2$ of the fit is 86, which corresponds
to a confidence level of 5.7\%, indicating a reasonable fit.
The main contributor to large $\chi^2$ is
DELPHI~\cite{23DELPHI:2003kmy}, the uncertainties of its three
measurements may have been underestimated, but there is no
significant effect on the results by including these data.

\begin{figure}[htbp]
\centering
  \includegraphics[width=0.45\textwidth]{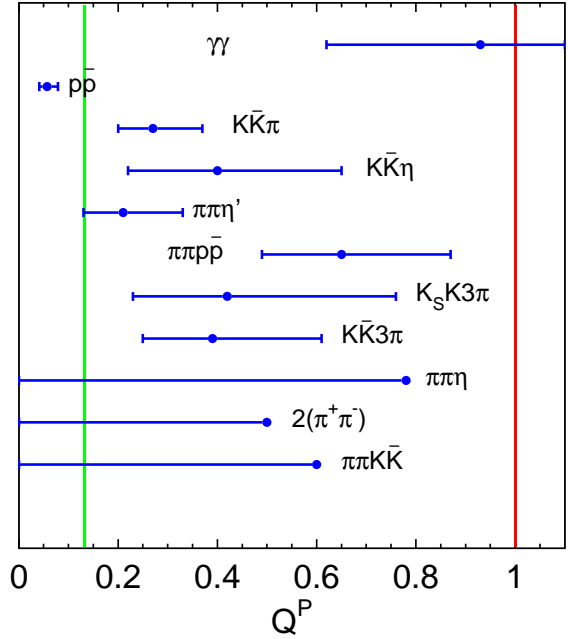}
\caption{$Q^P$ from the global fit and the comparison with
theoretical predictions. Dots with error bars are data and the
vertical lines show $Q^P=0.133$ and $Q^P=1$. } \label{fig:qp}
\end{figure}

Table~\ref{tab:results_1} and Figure~\ref{fig:qp} show the fit
results and the total uncertainties of $\etac$ and $\etacp$
hadronic decays. We can find that the ratios of all the modes with
positive $\etacp$ signals (upper half of
Table~\ref{tab:results_1}) are less than one, and those of some
modes with stringent $\etacp$ decay rates are also less than one,
although those of some other modes are inconclusive (lower half of
Table~\ref{tab:results_1}). These put the prediction of
$Q^P\approx 1$ in question. Although each and all of the ratios
agree with the ``12\% rule'' better than $Q^P\approx 1$, we can
find that all central values are higher than 13\% except for
$\ppb$ mode, which is lower by more than three standard
deviations. These indicate that the experimental measurements do
not agree with either of the two predictions.

\begin{table*}
 \centering
\caption{The branching fractions of $\etacp$ and $\etac$ decays
and the ratios. For the modes with upper limits only, the data are
from experimental measurements directly, the upper limits of the
ratios at the 90\% confidence level are determined by dividing the
upper limits of the $\etacp$ decays by the branching fractions of
the $\etac$ decays lowered by corresponding uncertainties.}
\label{tab:results_1}
\begin{tabular}{lcccc}
\hline\hline
 decay mode ($h$) & $\BR(\etac\to h)$ (\%) & $\BR(\etacp\to h)$ (\%) & $Q^P_h$  \\ \hline
 $p\bar{p}$       & $0.136\pm 0.012$          & $0.0077_{-0.0021}^{+0.0028}$  & $0.057_{-0.016}^{+0.022}$ \\
 $K\bar{K}\pi$    & $6.90_{-0.42}^{+0.44}$    & $1.86_{-0.49}^{+0.68}$        & $0.27_{-0.07}^{+0.10}$ \\
 $K\bar{K}\eta$   & $1.27_{-0.14}^{+0.15}$    & $0.51_{-0.23}^{+0.31}$        & $0.40_{-0.18}^{+0.25}$ \\
 $\pp\etap$       & $1.20_{-0.17}^{+0.18}$    & $0.25_{-0.09}^{+0.14}$        & $0.21_{-0.08}^{+0.12}$ \\
 $\pp\ppb$        & $0.365_{-0.039}^{+0.042}$ & $0.236_{-0.052}^{+0.076}$     & $0.65_{-0.16}^{+0.22}$ \\
 $\ks K^\pm\pi^\mp\pp$
                  & $2.39_{-0.62}^{+0.67}$    & $1.00_{-0.42}^{+0.69}$        & $0.42_{-0.19}^{+0.34}$\\
 $\kk\ppp$        & $3.50_{-0.57}^{+0.60}$    & $1.36_{-0.48}^{+0.70}$        & $0.39_{-0.14}^{+0.22}$\\ \hline
 $\pp\eta$        & $1.43^{+0.41}_{-0.38}$    & $<0.96$~\cite{CLEO:2009tno}   & $ <0.78$ \\
 $2(\pp)$         & $0.86^{+0.13}_{-0.12}$    & $<0.41$~\cite{Belle:2007qae}  & $ <0.50$ \\
 $\kk\pp$         & $0.57\pm 0.10$            & $<0.32$~\cite{Belle:2007qae}  & $ <0.60$ \\
 $2(\kk)$         & $0.135^{+0.028}_{-0.027}$ & $<0.14$~\cite{Belle:2007qae}  & $ <1.5 $ \\
 $3(\pp)$         & $1.75\pm 0.48$~\cite{BESIII:2012urf}
                                              & $<2.9$~\cite{CLEO:2009tno}    & $ <2.0 $ \\
 $\kk 2(\pp)$     & $0.72\pm 0.37$~\cite{BESIII:2012urf}
                                              & $<2.2$~\cite{CLEO:2009tno}    & $ <5.4 $\\
 $\phi\phi$       & $0.155^{+0.018}_{-0.017}$ & ---     & --- \\
 $\phi\kk$        & $0.36^{+0.15}_{-0.14}$    & ---     & --- \\
 $2(\ppp)$        & $15.1^{+2.0}_{-1.9}$      & ---     & --- \\
\hline\hline
\end{tabular}
\end{table*}

Table~\ref{tab:results_2} shows the other fit results. We can find
that the total widths of $\etac$ and $\etacp$ agree with those
obtained in Ref.~\cite{PDG}, and the ratio
  \( \Gamma_{\etacp}/\Gamma_{\etac} = 0.44\pm 0.10  \)
agrees well with the expectation of Eq.~(14) of
Ref.~\cite{Chao:1996sf}, i.e.,
\[
           \frac{\Gamma_{\etacp}}{\Gamma_{\etac}}
   \approx \frac{\Gamma_{\etacp\to hadrons}}{\Gamma_{\etac\to hadrons}}
   \approx \frac{\Gamma_{\psp\to \EE}}{\Gamma_{\jpsi\to \EE}}
   = 0.42 \pm 0.01,
\]
where $\Gamma_{\psp\to \EE}$ and $\Gamma_{\jpsi\to \EE}$ are the
leptonic partial widths of the vector charmonium
states~\cite{PDG}.

\begin{table}
 \centering
\caption{The fit results on $\etacp$ and $\etac$ related
quantities. } \label{tab:results_2}
\begin{tabular}{lc}
\hline\hline
 $\Gamma_{\etac}$    &   $32.2\pm 0.7$~MeV  \\
 $\Gamma_{\etacp}$   &   $14.1\pm 3.1$~MeV  \\\hline
 $\Gamma_{\etac\to \GG}$  &   $5.43_{-0.38}^{+0.41}$~keV \\
 $\Gamma_{\etacp\to \GG}$ &   $2.21_{-0.64}^{+0.88}$~keV \\\hline
 $\BR(\eth)$         &   $(10.8\pm 0.6)\times 10^{-4}$ \\
 $\BR(\epth)$        &   $(4.42\pm 0.96)\times 10^{-4}$ \\ \hline
 $\BR(\psp\to \gamma\etacp)$   & $(7.0^{+3.4}_{-2.5})\times 10^{-4}$ \\
 $\BR(\psp\to \piz\hc\to \piz\gamma\etac)$
                               & $(5.03^{+0.52}_{-0.49})\times 10^{-4}$ \\
 \hline\hline
\end{tabular}
\end{table}

The partial width of $\etac\to \GG$, $(5.43_{-0.38}^{+0.41})$~keV,
is about one standard deviation higher than the world
average~\cite{PDG} and is lower than the lattice QCD calculation
of $\Gamma_{\etac\to \GG}=(6.51\pm 0.20)$~keV~\cite{Meng:2021ecs}
by $2.5$ standard deviations. Further measurements and refined
calculations are needed to clarify the tension and to develop
other model
calculations~\cite{Brambilla:2018tyu,Chaturvedi:2019usm}.

The partial width of $\etacp\to \GG$ of
$(2.21_{-0.64}^{+0.88})$~keV is a first model-independent
evaluation, to be compared with various calculations compiled in
Ref.~\cite{Chaturvedi:2019usm}. It is worth to point out that the
ratio of the branching fractions of $\etacp$ and $\etac\to \GG$
agrees fairly well with the $Q^P=1$ rule:
\[
  \frac{\BR(\etacp\to \GG)}{\BR(\etac\to \GG)}
  = \frac{\Gamma_{\etacp\to \GG}/\Gamma_{\etacp}}{\Gamma_{\etac\to \GG}/\Gamma_{\etac}}
  = 0.93^{+0.48}_{-0.31},
\]
although the uncertainty is large.

As by products, we also report the world best evaluation of
$\BR(\eth)$, $\BR(\epth)$, $\BR(\psp\to \gamma\etacp)$, and
$\BR(\psp\to \piz\hc)\cdot \BR(\hc\to \gamma\etac)$, as shown in
Table~\ref{tab:results_2}. These results will be used in future
measurements with these processes.

In summary, we determined the ratios of the pseudoscalar
charmonium states $\etacp$ and $\etac$ decay branching fractions
and found prominent discrepancy from theoretical
predictions~\cite{Anselmino:1991es,Chao:1996sf}. The mixing of the
$\jpsi$ with a nearby glueball has been proposed~\cite{Hou:1982kh}
to explain the ``$\rho\pi$ puzzle'' and the scheme has been
extended to the $\etac$ case~\cite{Anselmino:1991es,Chao:1996sf}.
As the pseudoscalar glueballs are expected to be close to $\etac$
or $\etacp$~\cite{Huber:2021zqk,Chen:2021dvn}, the mixing between
them may also play important role in the charmonium
decays~\cite{Tsai:2011dp,Qin:2017qes,Zhang:2021xvl}. The different
contribution of the open-charm loop in $\etac$ and $\etacp$ decays
may affect the branching fraction ratio~\cite{Wang:2012wj} too.
The fact that all the known hadronic decays of $\etacp$ have rates
lower than $\etac$ decays suggests abnormal dynamics in either
$\etacp$ or $\etac$ decays, and these may be investigated at
future experiments like BESIII~\cite{bes3_white}, Belle
II~\cite{belle2}, and LHCb~\cite{lhcb_detector} in charmonium
decays, two-photon processes, and $B$ decays.


This work is supported in part by National Key Research and
Development Program of China under Contract No.~2020YFA0406301,
National Natural Science Foundation of China (NSFC) under contract
Nos. 11961141012, 11835012, and 11521505.


\appendixpage
\renewcommand{\baselinestretch}{1.1}

\begin{appendix}

\section*{Data used in the analysis
(Tables~\ref{tab:widgets},~\ref{tab:widgets2},~\ref{tab:widgets3},~\ref{tab:widgets4})}
\label{app}

\begin{table*}[hp]
\caption{Data used in the analysis: absolute branching fractions
and the ratios of the branching fractions for $\etac$ and
$\etacp$.  } \label{tab:widgets}
 \centering
\begin{tabular}{cllc}
\hline\hline
Index & quantity  & Value & Experiment \\\hline\
& Branching fraction & (\%) & \\ \hline
 \myct& $\BR(\efia)$ & $1.15 \pm 0.12 \pm 0.10$  & \bt~\cite{1BESIII:2019eyx} \\
 \myct& $\BR(\efib)$ & $2.60 \pm 0.21 \pm 0.20$  & \bt~\cite{1BESIII:2019eyx} \\
 \myct& $\BR(\eft)$  & $0.120\pm 0.026\pm 0.015$ & \bt~\cite{1BESIII:2019eyx} \\
 \myct& $\BR(\eset)$ & $15.3 \pm 1.8  \pm 1.8$   & \bt~\cite{1BESIII:2019eyx} \\
 \myct& $\BR(\eth)$  & $0.120\pm 0.008\pm 0.007$ & \be~\cite{30Belle:2017psv} \\
 \myct& $\BR(\eth)$  & $0.096\pm 0.012\pm 0.006$ & \bb~\cite{31BaBar:2019hzd} \\
 \myct& $\BR(\epth)$ & $0.048\pm 0.011\pm 0.003$ & \be~\cite{30Belle:2017psv} \\
 \myct& $\BR(\epth)$ & $0.035\pm 0.017\pm 0.005$ & \bb~\cite{31BaBar:2019hzd} \\ \hline
& Ratio of the branching fractions & & \\ \hline
 \myct& $\frac{\BR(\eel)}{\BR(\eft)}$ & $1.79\pm 0.14\pm 0.32$ & \lh~\cite{5LHCb:2017ymo} \\
 \myct& $\frac{\BR(\eel)}{\BR(\efi)}$ & $0.032^{+0.014}_{-0.010}\pm 0.009$ & \be~\cite{26Belle:2003oac}  \\
 \myct& $\frac{\BR(\es)}{\BR(\efia)}$ & $0.571\pm 0.025\pm 0.051$ & \bb~\cite{7BaBar:2014asx} \\
 \myct& $\frac{\BR(\etw)}{\BR(\efi)}$ & $0.052^{+0.016}_{-0.014}\pm 0.014$ & \be~\cite{26Belle:2003oac}  \\
 \myct& $\frac{\BR(\esit)}{\BR(\efi))}$ & $0.026^{+0.009}_{-0.007}\pm 0.007$ & \be~\cite{26Belle:2003oac} \\
 \myct& $\frac{\BR(\eps)}{\BR(\epfia)}$ & $0.82\pm 0.21\pm 0.27$& \be~\cite{7BaBar:2014asx} \\
 \myct& $\frac{\BR(\epfi)\cdot \BR(\epth)}{\BR(\efi)\cdot \BR(\eth)}$ & $0.096^{+0.020}_{-0.019}\pm0.025$ & \bb~\cite{16BaBar:2008fle} \\
 \myct& $\frac{\BR(\epfib)\cdot\Gamma_{\etacp\to \GG}}{\BR(\efib)\cdot\Gamma_{\etac\to \GG}}$ & $0.18\pm 0.05 \pm 0.02$ & CLEO~\cite{21CLEO:2003gwz} \\\hline\hline
\end{tabular}
\end{table*}

\begin{table*}[p]
\centering
\caption{Data used in the analysis: product branching fractions measured in $B$ decays
and charmonium decays.}
\label{tab:widgets2}
\begin{tabular}{clll}
\hline\hline
Index & quantity & Value ($\times 10^{-6}$) & Experiment \\
\hline
 \myct& $\BR(\eft)\cdot \BR(\pse)$ & $0.65\pm 0.19\pm 0.10$    & \bt~\cite{11BESIII:2012urf} \\
 \myct& $\BR(\eft)\cdot \BR(\et)$  & $0.224^{+0.038}_{-0.037}\pm 0.020$ & E835~\cite{24FermilabE835:2003ula} \\
 \myct& $ \BR(\eft)\cdot \BR(\et)$ & $0.336^{+0.080}_{-0.070}$ & E760~\cite{28E760:1995rep}\\
 \myct& $\BR(\eft)\cdot \BR(\eth)$ & $1.64\pm 0.41^{+0.17}_{-0.24}$ & \be~\cite{25Belle:2002jvr}\\
 \myct& $\BR(\eft)\cdot \BR(\ethb)$ & $1.79\pm 0.68^{+0.19}_{-0.25}$ & \be~\cite{25Belle:2002jvr}\\
 \myct& $\BR(\eft)\cdot \BR(\eth)$ & $1.8^{+0.3}_{-0.2}\pm 0.2$ & \bb~\cite{32BaBar:2005sdl}\\
 \myct& $\BR(\eft)\cdot \BR(\eth)$ & $1.42\pm 0.11^{+0.16}_{-0.20}$ & \be~\cite{19Belle:2006vmn}\\
 \myct& $\BR(\et)\cdot \BR(\eth)$  & $0.22^{+0.09}_{-0.07}\,^{+0.04}_{-0.02}$ & \be~\cite{18Belle:2006mzv}\\
 \myct& $\BR(\eel)\cdot \BR(\eth)$ & $4.7\pm 1.2\pm 0.5$ & \bb~\cite{22BaBar:2004phb}\\
 \myct& $\BR(\eel)\cdot \BR(\ethb)$ & $2.4\pm 1.4\pm 0.3$ & \bb~\cite{22BaBar:2004phb}\\
 \myct& $\BR(\efi)\cdot \BR(\eth)$  & $74.0\pm 5.0\pm 7.0$ & \bb~\cite{22BaBar:2004phb} \\
 \myct& $\BR(\efi)\cdot \BR(\ethb)$ & $64.8\pm 8.5\pm 7.1$ & \bb~\cite{22BaBar:2004phb}\\
 \myct& $\BR(\efia)\cdot \BR(\pse)$ & $4.54\pm 0.76\pm 0.48$ & \bt~\cite{11BESIII:2012urf}\\
 \myct& $\BR(\efia)\cdot \BR(\eth)$ & $11.4\pm 2.5^{+1.1}_{-1.8}$ & \be~\cite{25Belle:2002jvr}\\
 \myct& $\BR(\efia)\cdot \BR(\ethb)$ & $16.6\pm 5.0\pm 1.8$ & \be~\cite{25Belle:2002jvr}\\
 \myct& $\BR(\efib)\cdot \BR(\pse)$  & $11.35\pm 1.25\pm 1.50$ & \bt~\cite{11BESIII:2012urf}\\
 \myct& $\BR(\efib)\cdot \BR(\eth)$  & $24.0\pm 1.2^{+2.1}_{-2.0}$ & \be~\cite{14Belle:2011mee}\\
 \myct& $\BR(\efib)\cdot \BR(\ethb)$ & $20.1\pm 4.7^{+3.0}_{-4.5}$ & \be~\cite{25Belle:2002jvr}\\
 \myct& $\BR(\etht)\cdot \BR(\pse)$  & $7.22\pm 1.47\pm 1.11$ & \bt~\cite{11BESIII:2012urf}\\
 \myct& $\BR(\es)\cdot \BR(\pse)$    & $2.11\pm 1.01\pm 0.32$ & \bt~\cite{11BESIII:2012urf}\\
 \myct& $\BR(\ef)\cdot \BR(\pse)$     & $7.51\pm 0.85\pm 1.11$ & \bt~\cite{11BESIII:2012urf}\\
 \myct& $\BR(\esit)\cdot \BR(\pse)$   & $0.94\pm 0.37\pm 0.14$ & \bt~\cite{11BESIII:2012urf}\\
 \myct& $\BR(\esit)\cdot \BR(\eth)$   & $2.0\pm 0.6\pm 0.4$ & \bb~\cite{22BaBar:2004phb}\\
 \myct& $\BR(\esit)\cdot \BR(\ethb)$  & $0.9\pm 0.9\pm 0.4$ & \bb~\cite{22BaBar:2004phb}\\
 \myct& $\BR(\ee)\cdot \BR(\eth)$   & $3.94^{+0.41}_{-0.39}\,^{+0.22}_{-0.18}$ & \be~\cite{2Belle:2019avj}\\
 \myct& $\BR(\ee)\cdot \BR(\etha)$  & $1.90^{+0.32}_{-0.29}\,^{+0.13}_{-0.47}$ & \be~\cite{2Belle:2019avj}\\
 \myct& $\BR(\ee)\cdot \BR(\pse)$   & $2.30\pm 0.65\pm 0.36$ & \bt~\cite{11BESIII:2012urf}\\
 \myct& $\BR(\efit)\cdot \BR(\pse)$ & $4.16\pm 0.76\pm 0.59$ & \bt~\cite{11BESIII:2012urf}\\
 \myct& $\BR(\en)\cdot \BR(\pse)$   & $12.01\pm 2.22\pm 2.04$ & \bt~\cite{11BESIII:2012urf}\\
 \myct& $\BR(\eset)\cdot \BR(\pse)$ & $75.13\pm 7.42\pm 9.99$ & \bt~\cite{11BESIII:2012urf}\\
 \myct& $\BR(\epf)\cdot \BR(\epth)$ & $0.0342\pm 0.0071\pm 0.0021$ & \lh~\cite{4LHCb:2016zqv} \\
 \myct& $\BR(\epfi)\cdot \BR(\psep)$ & $13.0\pm 2.0\pm 3.0$ & \bt~\cite{10BES:2012uhz}\\
 \myct& $\BR(\epfib)\cdot \BR(\epth)$ & $3.1\pm 0.8\pm 0.2$ & \be~\cite{14Belle:2011mee}\\
 \myct& $\BR(\epe)\cdot \BR(\epth)$  & $1.12^{+0.18}_{-0.16}\ ^{+0.05}_{-0.07}$ & \be~\cite{2Belle:2019avj}\\
 \myct& $\BR(\epe)\cdot \BR(\eptha)$ & $0.42^{+0.14}_{-0.12}\pm 0.03$ & \be~\cite{2Belle:2019avj} \\
 \myct& $\BR(\epn)\cdot \BR(\psep)$ & $7.03\pm 2.10\pm 0.70$ & \bt~\cite{8BESIII:2013nja}\\
\hline\hline
\end{tabular}
\end{table*}

\begin{table*}[p]
\centering \caption{Data used in the analysis: product of $\GG$
partial width and branching fraction of $\etac$ and $\etacp$
decays measured in two-photon processes. } \label{tab:widgets3}
\begin{tabular}{clll}
\hline\hline
Index & quantity & Value (eV) & Experiment \\ \hline
 \myct& $\BR(\eel)\cdot\Gamma_{\etac\to \GG}$ & $7.75\pm 0.66\pm 0.62$ & \be~\cite{12Belle:2012qqr} \\
 \myct& $\BR(\eel)\cdot\Gamma_{\etac\to \GG}$ & $6.8\pm 1.2\pm 1.3$ &  \be~\cite{Belle:2007qae} \\
 \myct& $\BR(\eft)\cdot\Gamma_{\etac\to \GG}$ & $7.20\pm 1.53^{+0.67}_{-0.75}$ & \be~\cite{20Belle:2005fji} \\
 \myct& $\BR(\efi)\cdot\Gamma_{\etac\to \GG}$ & $386\pm 8\pm 21$ & \bb~\cite{13BaBar:2011gos} \\
 \myct& $\BR(\efi)\cdot\Gamma_{\etac\to \GG}$ & $374\pm 9\pm 31$ & \bb~\cite{15BaBar:2010siw} \\
 \myct& $\BR(\efi)\cdot\Gamma_{\etac\to \GG}$ & $600\pm 120\pm 90$ &   DELPHI~\cite{23DELPHI:2003kmy} \\
 \myct& $\BR(\efib)\cdot\Gamma_{\etac\to \GG}$& $490\pm 290\pm 90$ & AMY~\cite{27AMY:1998ghf} \\
 \myct& $\BR(\efib)\cdot\Gamma_{\etac\to \GG}$& $142\pm 4\pm 14$ & \be~\cite{29Nakazawa:2008zz} \\
 \myct& $\BR(\ese)\cdot\Gamma_{\etac\to \GG}$ & $65.4\pm 2.6\pm 7.8$ & \be~\cite{3Belle:2018bry} \\
 \myct& $\BR(\ef)\cdot\Gamma_{\etac\to \GG}$  & $40.7\pm 3.7\pm 5.3$ & \be~\cite{Belle:2007qae} \\
 \myct& $\BR(\efit)\cdot\Gamma_{\etac\to \GG}$& $25.7\pm 3.2\pm 4.9$ & \be~\cite{Belle:2007qae} \\
 \myct& $\BR(\efit)\cdot\Gamma_{\etac\to \GG}$& $280\pm 100\pm 60$ & DELPHI~\cite{23DELPHI:2003kmy} \\
 \myct& $\BR(\esit)\cdot\Gamma_{\etac\to \GG}$& $5.6\pm 1.1\pm 1.6$ & \be~\cite{Belle:2007qae} \\
 \myct& $\BR(\esit)\cdot\Gamma_{\etac\to \GG}$& $350\pm 90\pm 60$ & DELPHI~\cite{23DELPHI:2003kmy} \\
 \myct& $\BR(\ete)\cdot\Gamma_{\etac\to \GG}$ & $190\pm 6\pm 28$ & \bb~\cite{13BaBar:2011gos} \\
 \myct& $\BR(\epfi)\cdot\Gamma_{\etacp\to \GG}$ & $41\pm 4\pm 6$ & \bb~\cite{13BaBar:2011gos} \\
 \myct& $\BR(\epfib)\cdot\Gamma_{\etacp\to \GG}$ & $11.2\pm 2.4\pm 2.7$ & \be~\cite{29Nakazawa:2008zz} \\
 \myct& $\BR(\epse)\cdot\Gamma_{\etacp\to \GG}$ & $5.6^{+1.2}_{-1.1}\pm 1.1$ &   \be~\cite{3Belle:2018bry} \\
 \myct& $\BR(\epte)\cdot\Gamma_{\etacp\to \GG}$ & $30\pm 6\pm 5$ & \bb~\cite{13BaBar:2011gos} \\
\hline\hline
\end{tabular}
\end{table*}

\begin{table*}[p]
\caption{Data used in the analysis: total widths of $\etac$ (upper
half in the table) and $\etacp$ (lower half in the table).}
\label{tab:widgets4} \centering
\begin{tabular}{clll}
\hline\hline
Index & Process & Width (MeV) & Experiment \\ \hline
\myct& $\GG\to \etac$, $\etac\to \etap\pp$ & $30.8^{+2.3}_{-2.2}\pm 2.9$ & \be~\cite{3Belle:2018bry} \\
\myct& $\GG\to \etac$, $\etac\to \kk\eta$  & $34.8\pm 3.1\pm 4.0$ & \bb~\cite{7BaBar:2014asx} \\
\myct& $\GG\to \etac$, $\etac\to \kk\piz$  & $25.2\pm 2.6\pm 2.4$ & \bb~\cite{7BaBar:2014asx} \\
\myct& $\GG\to \etac$, $\efib$ & $32.1\pm 1.1\pm 1.3$ & \bb~\cite{13BaBar:2011gos} \\
\myct& $\GG\to \etac$, $\efib$ & $24.8\pm 3.4\pm 3.5$ & \cl~\cite{21CLEO:2003gwz} \\
\myct& $\GG\to \etac$, $\efib$ & $36.6\pm 1.5\pm 2.0$ & \be~\cite{29Nakazawa:2008zz} \\
\myct& $\GG^{*}\to \etac$, $\efib$ & $31.7\pm 1.2\pm 0.8$ & \bb~\cite{15BaBar:2010siw} \\
\myct& $\GG\to \etac$, $\ete$      & $36.2\pm 2.8\pm 3.0$ & \bb~\cite{13BaBar:2011gos} \\
\myct& $\GG\to \etac$, $\etac\to hadrons$  & $28.1\pm 3.2\pm 2.2$ & \be~\cite{Belle:2007qae} \\
\myct& $B^+\to \etac K^+$, $\etac\to \ppb$ & $34.0\pm 1.9\pm 1.3$ & \lh~\cite{4LHCb:2016zqv} \\
\myct& $B^+\to \etac K^+$, $\eft$          & $48^{+8}_{-7}\pm 5$ & \be~\cite{19Belle:2006vmn} \\
\myct& $B^+\to \etac K^+$, $\etac\to \Lambda\bar{\Lambda}$ & $40\pm 19\pm 5$ & \be~\cite{19Belle:2006vmn} \\
\myct& $B^+\to \etac K^+$, $\efib$         & $35.4\pm 3.6^{+3.0}_{-2.1}$ & \be~\cite{14Belle:2011mee} \\
\myct& $B\to \etac K^{(*)}$, $\efi$        & $36.3^{+3.7}_{-3.6}\pm 4.4$ & \bb~\cite{16BaBar:2008fle} \\
\myct& $b\to \etac X$, $\etac\to \phi\phi$ & $31.4\pm 3.5\pm 2.0$ & \lh~\cite{5LHCb:2017ymo} \\
\myct& $b\to \etac X$, $\etac\to \ppb$     & $25.8\pm 5.2\pm 1.9$ & \lh~\cite{6LHCb:2014oii} \\
\myct& $p\bar{p}\to\etac$, $\etac\to\GG$ & $20.4^{+7.7}_{-6.7}\pm 2.0$ & E835~\cite{24FermilabE835:2003ula} \\
\myct& $p\bar{p}\to\etac$, $\etac\to\GG$ & $23.9^{+12.6}_{-7.1}$ & E760~\cite{28E760:1995rep} \\
\myct& $\pse$, $\etac\to hadrons$ & $32.0\pm 1.2\pm 1.0$ & \bt~\cite{9BESIII:2011ab} \\
\myct& $\pse$, $\etac\to hadrons$ & $36.4\pm 3.2\pm 1.7$ & \bt~\cite{9BESIII:2011ab} \\
\hline
\myct& $\GG\to \etacp$, $\epfib$         & $13.4\pm 4.6\pm 3.2$ & \bb~\cite{13BaBar:2011gos} \\
\myct& $\GG\to \etacp$, $\epfib$ & $6.3\pm 12.4\pm 4.0$ & \cl~\cite{21CLEO:2003gwz} \\
\myct& $\GG\to \etacp$, $\epfib$ & $19.1\pm 6.9\pm 6.0$ & \be~\cite{29Nakazawa:2008zz} \\
\myct& $B^+\to \etacp K^+$, $\epfib$     & $41.0\pm 12.0^{+6.4}_{-10.9}$ & \be~\cite{14Belle:2011mee} \\
\myct& $\psp\to \gamma \etacp$, $\epfi$ & $16.9\pm 6.4\pm 4.8$ & \bt~\cite{10BES:2012uhz} \\
\myct& $\psp\to \gamma \etacp$, $\epn$  & $9.9\pm 4.8\pm 2.9$ & \bt~\cite{8BESIII:2013nja} \\
\hline\hline
\end{tabular}
\end{table*}

\end{appendix}

\end{document}